\documentclass[twocolumn,showpacs,preprintnumbers,amsmath,amssymb,prb]{revtex4}

\usepackage{graphicx}
\usepackage{dcolumn}
\usepackage{bm}

\begin{document}
\title{Experimental measurement and Finite element modeling of extraordinary optoconductance in GaAs-In metal-semiconductor hybrid structures}
\author{K. A. Wieland}
\author{Yun Wang}
\author{S. A. Solin}
\email{solin@wustl.edu}
\affiliation{Department of Physics and Center for Materials Innovation, Washington University in St. Louis, CB 1105, 1
Brookings Drive, St. Louis, MO 63130}

\author{A. M. Girgis}
\author{L. R. Ram-Mohan}
\affiliation{Department of Physics, Worcester Polytechnic Institute, Worcester, MA 01609}%

\date{March 7, 2006}

\begin{abstract} We present a detailed discussion of extraordinary optoconductance (EOC).   [Here optoconductance is synonymous with photoconductance.] Experimental data
was acquired via macroscopic metal-semiconductor hybrid structures composed of GaAs and In
and subjected to illumination from an Ar ion laser. A drift diffusion model using the finite element
method (FEM) provided a reasonable fit to the data. EOC is explored as a function of laser position, 
bias current, laser power density, and temperature. The positional dependence of the voltage
is accounted for by the Dember effect, with the model incorporating the excess hole distribution
based on the carrier mobility, and thus the mean free path.  The bias current is found to 
produce a linear voltage offset and does not influence the EOC.  A linear relationship is found 
between the laser power density and the voltage in the bare and hybrid devices. This dependence
is reproduced in the model by a generation rate parameter which is related to the power density.
Incorporating the mobility and diffusion temperature dependence, the model directly parallels
the temperature dependence of the EOC without the use of fitting parameters. 
\end{abstract}

\maketitle

\section{\label{sec:intro}INTRODUCTION}

The resistance of any device can be divided into two components, one physical and one geometric. Contributions to the
physical component arise from factors such as the doping levels, impurities, bulk conductivity, and bulk mobility. The
geometric contributions arise from factors such as the device dimensions, shape, and arrangement of the leads as well as
any inhomogeneities that may be present. Most often, the physical contributions dominate the resistance, but by
judicious choices, the geometric component can become the dominant component. The recently discovered ÒEXXÓ
effects\cite{1} 
in metal-semiconductor hybrid structures (MSHs) of the type shown schematically in Fig. 1(b) are primarily geometric in
nature and rely on a reallocation of current between the metal and semiconductor regions due to an external
perturbation. Here XX refers to magnetoresistance (MR), piezoconductance (PC), etc. The perturbation can alter the
interface between the metal and semiconductor regions or, as we show here, the bulk transport characteristics of either
constituent. If the metal and semiconductor conductivities differ by several orders of magnitude, as is the case with
GaAs and In, small perturbations can dramatically change the flow of current through the device.  As a result, under
perturbation, the measured voltage difference (depicted as V$_MSH$ in Fig. 1(b)) increases as compared to that of the metal or
semiconductor alone. The large, often linear, change in voltage makes EXX devices an excellent choice for high
resolution spatial sensing. Additionally, because they are non-magnetic, EXX devices can be used in applications where
typical magnetic sensors are not suitable.\cite{2}

The original EXX effect to be discovered was EMR.\cite{3}   An InSb-Au MSH structure of macroscopic (and later nanometer)
dimensions was fabricated using novel techniques described elsewhere.\cite{4}  The perturbation in this case was a
magnetic field, redirecting the current through the semiconductor, whereas that current flows primarily through the
metal with no applied field.  A room temperature EMR of order $10^{6}$\% was observed at fields of 5T. Theoretical
modeling of EMR, using the finite element method (FEM) incorporating the effects of the boundary conditions at the
metal-semiconductor interface, has shown excellent agreement with the experimental data.\cite{5,6} EMR has potential
applications in a number of diverse areas including ultra-high-density magnetic recording, position sensing, and medical
instrumentation.  Scanning EMR probes have also been proposed as a way to concurrently measure topographic and magnetic
field information with high spatial and temporal resolution.\cite{7} 

The second demonstrated EXX effect was EPC.\cite{8}   An InSb-Au MSH structure, similar to that used for EMR, was constructed. 
In EPC the perturbation is a uniaxial tensile strain of the metal-semiconductor interface.  Room temperature
extraordinary piezoconductance as high as $500$\% has been demonstrated in the above described MSH structures.\cite{8}
Theoretical models based on both the FEM solution and the analytic solutions of LaplaceÕs equation have accurately
reproduced the EPC effect.\cite{1} EPC devices have obvious applications in the construction industry and have also been
proposed for studying basic material properties (interface dynamics). 

Following the discovery of EMR and EPC it was realized that an optical equivalent of the EXX phenomena, e.g.
extraordinary optoconductance or EOC, should exist, with photons providing the perturbation.  Such a device would have
applicability in the solar energy field, modern fiber optic relays, position sensors, and other optoelectronic devices. 
We have recently provided a proof of principle demonstration of the EOC effect in GaAs-In MSH structures in a brief
preliminary report.\cite{9}   Here we give a full account of this effect and show a detailed FEM calculation that
quantitatively accounts for the observed experimental results. 

A diagram of an MSH device is given in Fig. 1(b) with the
connections shown schematically. The mesh shown in Fig. 1(c) contains fewer elements than the actual mesh used for
calculations. The bare device has the same physical connections as shown in Fig. 1(a). In a voltage measurement,
the surface of the semiconductor region is exposed to a focused laser beam, whose lateral position $(x_{\ell},y_{\ell})$
can be accurately controlled (see experimental details below).  We define EOC as the percent difference in the measured
output voltage in the MSH, as compared to that of the bare semiconductor with no shunt attached, such that
\begin{eqnarray}
EOC(x_{\ell},y_{\ell},\lambda,\{\beta\})=\nonumber\\
\left(\frac{[V_{23}(x_{\ell},y_{\ell})]_{MSH}-[V_{23}(x_{\ell},y_{\ell})]_{bare}}{[V_{23}(x_{\ell},y_{\ell})]_{bare}}\right)\times
100\%.
\end{eqnarray}
Here the parameter set \{$\beta$\} represents the power density of the laser, $P$, the temperature, $T$, and the bias
current $I$. The quantity $\lambda=476.5$ nm is the emission wavelength of the laser, which is held fixed for all
experiments.
In this paper, we investigate the EOC as a function of the laser spot position with the parameters \{$\beta$\} held
fixed, and as a function of each of the \{$\beta$\} parameters with the spot position and the other parameters in the
set held fixed. Because of the many factors influencing EOC, it presents a unique opportunity to study the dynamics of
the metal semiconductor interface, carrier transport phenomena, and the effect of introducing a shunt.
\begin{figure}
\includegraphics[width=0.5\textwidth]{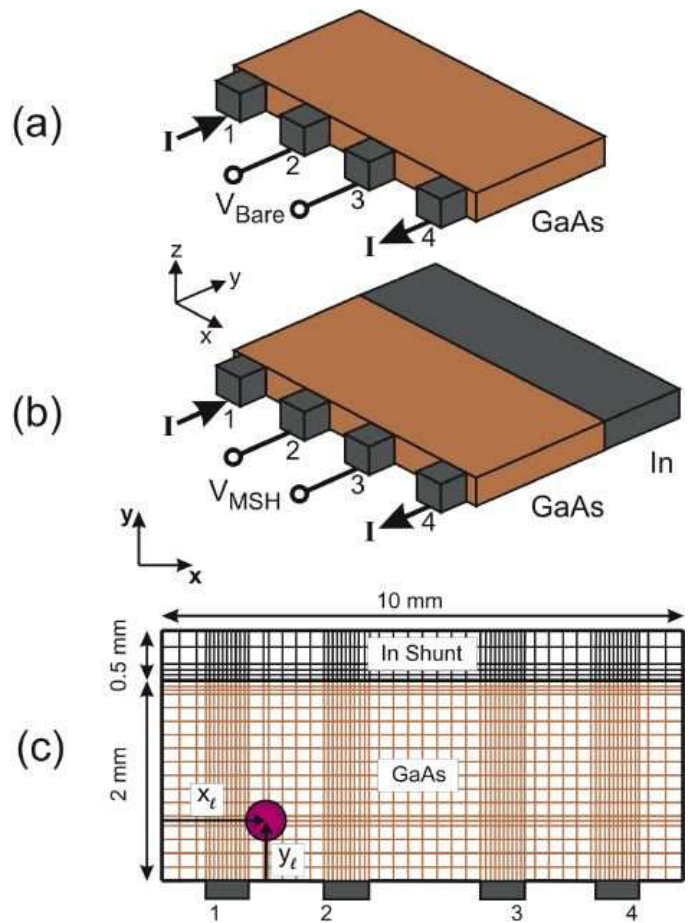}
\caption{\label{fig:FEMmesh}(Color online) Bare (a) and MSHs (b) devices.  (c) The FEM mesh showing the device dimensions, port configuration, and schematic of the connections to the device.}
\end{figure}
\section{\label{sec:experiment}EXPERIMENT}
\subsection{\label{sec:expsetup}Experimental Setup} The experimental setup used to measure the EOC is shown in Figure 2.
The bare (shuntless) sample was illuminated with a Coherent Innova 400 Ar+ ion laser operating at $476.5$ nm with the
beam focused to a 20 $\mu$m diameter. The power density was varied from $6.3\times10^{4}$ W/cm$^{2}$ to $5.8\times10^{6}
$ W/cm$^{2}$. The samples were mounted in an ARS closed cycle helium cryostat and cooled to temperatures ranging from 10 K
- 300 K. The cryostat temperature was maintained by a Scientific Instruments 9650 temperature controller. The sample
was positioned using a Newport Universal Motion Controller ESP 300 system with three linear DC stepper motors. For
optimal spatial resolution, the sample was placed at the focused beam waist, thereby defining the distance from the lens
to the sample and the diameter of the laser spot. The bias current was supplied by a Lakeshore 120 current source in the
forward and reverse direction from 1 $\mu$ A to 100 mA across ports 1 and 4, as shown in Fig. 1.  The voltage was
acquired by a Keithley 2182 nanovoltmeter. One channel measured   and the other channel (not shown) measured the voltage
drop across the 1.2 $\Omega$ resister in series with the sample. The voltage across the resistor was used to precisely
determine the bias current flowing through the device.
\begin{figure}
\includegraphics[width=0.5\textwidth]{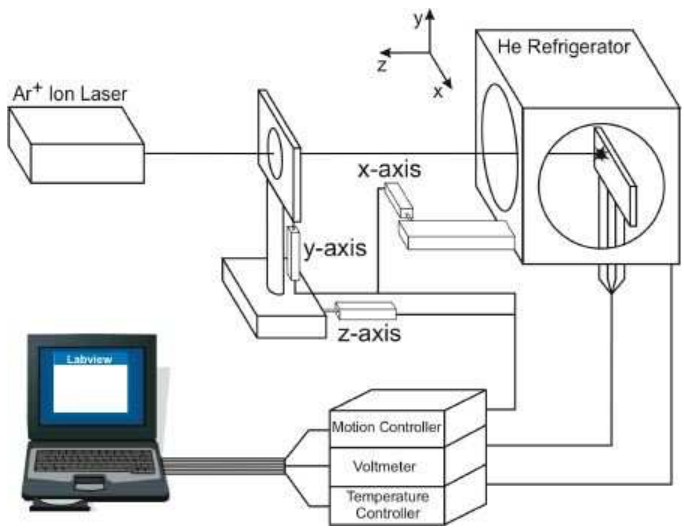}
\caption{\label{fig:schematic}(Color online) A schematic diagram of the experimental setup for sample positioning and data acquisition.}
\end{figure}
\subsection{\label{sec:samplechar}Sample Characterization} Bare samples of dimension 2 mm $\times$ 10 mm $\times$ 0.4 mm
were prepared by dicing a 2-inch diameter \textit{n}-type GaAs wafer with the [001] growth direction along the 10 mm
side. Gallium arsenide was chosen because it is a direct gap\cite{10,11} ($1.424$ eV at $300$ K) semiconductor and
because of optimal absorption in the spectral region of the argon ion laser.  The sample was degenerately doped\cite{12}
with Si at a concentration $N_{D} = 1.25\times10^{18}$cm$^{-3}$. The electron mobility of the semiconductor was
measured to
be $2100$ cm$^{2}/$Vs. Leads were attached first by metalizing the GaAs, in an inert nitrogen atmosphere, with dots of
In (of diameter $\sim0.2$ mm) arranged along the 10 mm side (see Fig. 2). 
Indium was chosen for metalizing because of its low melting temperature and compatibility with GaAs.\cite{13}   This
setup is similar to the typical four-lead van der Pauw plate setup used elsewhere.\cite{14}   This geometry is the conformal equivalent\cite{4} to a van der Pauw disk with an off-centered inclusion. Next, the wires were
tinned with In and pressed onto the metalized In dots.  Ohmic contacts over the temperature range of interest were
confirmed by verifying the linearity of the measured I-V response. These same bare samples were then used to make hybrid
samples via the addition of an In shunt.  This was achieved by metalizing the bare samples on the side opposite the
leads with Indium of dimensions 10 mm $\times$ 0.5 mm $\times$ 0.5 mm. The hybrid sample resistance at room temperature,
as measured across ports 1 and 4, decreased by an order of magnitude from that of the bare samples, indicating the
effectiveness of the shunt. Ohmic contact at the interface was again verified by the linearity of the I-V response.
\subsection{\label{sec:expproc}Experimental Procedure} Once the sample position was calibrated, the positional dependence
of the voltage $V_{23}$ was studied by moving the laser spot along the $x$-direction for fixed values of $y_{\ell}$.
However, in no case was the interface between the semiconductor and the shunt illuminated by the focused laser beam.
Voltages were read every 0.1 mm in the $x$-direction which provided adequate resolution. Upon reaching the end of the
sample, the $x$-position was returned to zero and the $y$-position was incremented by 0.1 mm. The program Labview was
used to control the sample position and to acquire the data. The experiments were performed under the same conditions
for the bare sample and the MSH in order to produce consistent results. The position-dependent data acquisition, as
outlined above, was repeated for various temperatures, bias currents, and power densities.
\section{\label{sec:model}THEORETICAL MODEL} When the laser illuminates a circular region of radius $r_{s}$, at a
position $(x_{\ell},y_{{\ell}})$, electron-hole pairs are created within the region and the individual carriers diffuse
away from the spot center. As noted by Dember\cite{15} many years ago, the electrons, which have a much higher mobility
than the holes $(\mu_{e}/\mu_{h}\sim20$ in GaAs$)$,\cite{16} diffuse away more rapidly from the focal spot region which
causes an excess steady state positive charge to develop there. Outside the focal region, the diffusing carriers
eventually recombine due to a number of processes. Nevertheless, the excess charge in the focal region perturbs the
potential at the voltage probes and this perturbation depends, not only on the position of the focal spot but also on a
number of other factors, such as temperature, excitation wavelength, power density, etc. Most importantly, the voltage
probe perturbation also depends on the proximity to the shunt that, if present, enhances the removal of the
photo-induced carriers and thus quantitatively affects the magnitude of the excess charge in the focal region. The
processes described above can be modeled using the semiconductor drift-diffusion equations with terms that account for
photo-generation and recombination. The electric field appearing in these equations is obtained from the solution of the
PoissonÕs equation with the excess carriers as the source terms. Because the skin depth (100 nm at a wavelength of 476.5 nm) and the hole diffusion length computed according to Chuang\cite{17} [$L_{h} = 0.1$ nm (see also discussion in section \textbf{III}. \textit{A}. below)] that  corresponds to several carrier lifetimes is a fraction of our sample thickness, the problem becomes two dimensional.
\subsection{\label{sec:driftdiff}The Drift-Diffusion Equations} Following the arguments of McKelvey\cite{17} and
Chuang,\cite{18} when photo-generated carriers are present, the current has the usual resistive and diffusive terms,
\begin{eqnarray}
\vec{J_{n}}&=&q(\mu _{n}n\vec{E}+D_{n}\vec{\nabla}n)\\
\vec{J_{p}}&=&q(\mu _{p}p\vec{E}-D_{p}\vec{\nabla}p),
\end{eqnarray}
where $n$ and $p$ represent the electron and hole number densities and $\mu_{n}$ and $\mu_{p}$ are their respective
mobilities.  The quantity $q=|e|$ is the magnitude of the electron charge.  The quantities $D_{n}$  and $D_{p}$  are the
diffusion coefficients defined by the Einstein relations,  
\begin{eqnarray}
D_{n}&=&\frac{\mu_{n}k_{b}T}{q}\\
D_{p}&=&\frac{\mu_{p}k_{b}T}{q}.
\end{eqnarray}
The effects of recombination and e-h pair generation must be taken into account in the equations of continuity so that
we have
\begin{eqnarray}
\frac{1}{q}\vec{\nabla}\cdotp\vec{J_{n}}+G(x,y)-R_{n}(x,y)&=& \frac{\partial n}{\partial t}\\
-\frac{1}{q}\vec{\nabla}\cdotp\vec{J_{p}}+G(x,y)-R_{p}(x,y)&=& \frac{\partial p}{\partial t},
\end{eqnarray}
where $G(x,y)$ is the spatially dependent generation rate per unit volume, and $R_{n}(x,y), R_{p}(x,y)$ are the
recombination rates for electrons and holes. 

Using Eqs. (2) and (3) in the above equations gives the steady state semiconductor drift-diffusion equations,
\begin{eqnarray}
\vec{\nabla}\cdotp(D_{n}\vec{\nabla}n+\mu _{n}n\vec{E})+G(x,y)-R_{n}(x,y)&=&0\\
\vec{\nabla}\cdotp(D_{p}\vec{\nabla}p-\mu _{p}p\vec{E})+G(x,y)-R_{p}(x,y)&=&0
\end{eqnarray}
The carrier number densities can be written as $n=n_{o}+\bar{n}$ and $p=p_{o}+\bar{p}$.  In the case at hand, the
materials are \textit{n}-type with $p_{o}\cong 0$, and the equilibrium electron number density $n_{o}$ is much larger
than the excess electron density $\bar{n}$.  As noted above, in GaAs the electron mobility is roughly 20 times larger
than that of holes.  As a result, the excess electrons $(\bar{n})$ diffuse away rapidly from the laser spot and therefore can be
viewed as a small increase in the large equilibrium electron density. 
Moreover, since $p_{o}\cong 0$, the hole number density is simply equal to the excess hole number density $\bar{p}$. From these assumptions follows an approximate charge neutrality, in which the charge density remains close to $n_{0}$. The drift-diffusion equation for holes can then be written as,
\begin{eqnarray}
\vec{\nabla}\cdotp(D_{p}\vec{\nabla}\bar{p}-\mu _{p}\bar{p}\vec{E})+G(x,y)-R_{p}(x,y)&=&0,
\end{eqnarray}
where the quantity $\vec{E}$ is the local electric field. The recombination rate for holes is assumed to be proportional
to the hole number density, $R_{p}=\frac{\bar{p}(x,y)}{\tau}$, where $\tau=0.5$ ns is the carrier recombination
time\cite{17} used in our calculations. The generation rate $G(x,y)$ is assumed to be a Gaussian, with the average
radius being equal to the spot radius of the illuminating laser
\begin{eqnarray}
G(x,y)&=& G_{o}exp\left[-\frac{\pi}{4}\left(\frac{r}{r_{s}}\right)^{2}\right].
\end{eqnarray}
Here $G_{o}$ is the generation rate per unit volume at the spot center, and the value of the coordinate $r$ is relative
to the spot center position $(x_{\ell},y_{\ell})$. The drift diffusion equation for holes is now given by
\begin{eqnarray}
\vec{\nabla}\cdotp(D_{p}\vec{\nabla}\bar{p}-\mu
_{p}\bar{p}\vec{E})+G_{o}exp\left[-\frac{\pi}{4}(\frac{r}{r_{s}})^{2}\right]-\frac{\bar{p}}{\tau}=0,
\end{eqnarray}
In order to solve Eq. (12), we set (a) $\bar{p}=0$ on the current ports 1 and 4, and (b) $\bar{p}=0$ in the metallic
shunt region. The first boundary condition asserts that any net hole distribution on the current ports will recombine
with the electrons in the Indium leads. The second boundary condition is due to the fact that the shunt is metallic and
therefore any hole states will be short lived. These boundary conditions are straightforward to implement within the
framework of the FEM. Figure 3 shows the hole distribution for both the bare and shunted device. The hole distribution
for the bare device is delocalized as compared to that of the shunted device. 
\begin{figure}
\includegraphics[width=0.5\textwidth]{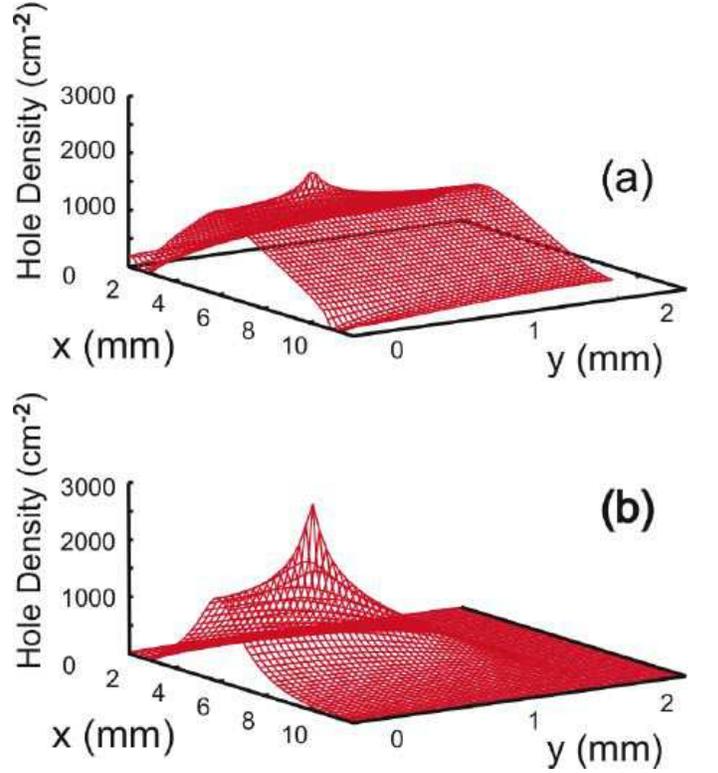}
\caption{\label{fig:holedens}(Color online) Excess hole density in (a) the bare device and (b) the MSH device at the laser spot position
$(x_{\ell},y_{\ell})=(4$mm$, 0.7$mm$)$.}
\end{figure}
\subsection{\label{sec:FEM}The Finite Element Method} 
The solution of Eq. (12) is obtained using the Galerkin
FEM,\cite{19,20}  with the weight functions being the same functions as the interpolation polynomials used to represent
the number density $\bar{p}$ in each finite element.  Multiplying Eq. (12) by the Galerkin weight function $\phi(x,y)$
and integrating gives
\begin{eqnarray}
\iint\phi\vec{\nabla}\cdotp(D_{p}\vec{\nabla}\bar{p}-\mu_{p}\bar{p}\vec{E})dA-
\iint\phi\frac{\bar{p}}{\tau}dA=\nonumber\\
-G_{o}\iint\phi
exp\left[-\frac{\pi}{4}\left(\frac{r}{r_{s}}\right)^{2}\right]dA.
\end{eqnarray}
The first integral in the above equation is integrated by parts to yield
\begin{eqnarray}
-\iint\vec{\nabla}\phi\cdotp(D_{p}\vec{\nabla}\bar{p}-\mu_{p}\bar{p}\vec{E})dA+\nonumber\\
\oint\phi(D_{p}\vec{\nabla}\bar{p}-\mu\bar{p}\vec{E})\cdotp\hat{n}dl-
\iint\phi\frac{\bar{p}}{\tau}dA=\nonumber\\
-G_{o}\iint\phi exp\left[-\frac{\pi}{4}\left(\frac{r}{r_{s}}\right)^{2}\right]dA.\nonumber
\end{eqnarray}
The second term, the ÒsurfaceÓ term, in the above equation represents a flux of holes through the external boundary and
is included only at the current ports in Fig. 1(c). The hole current at the current ports can be calculated using
\begin{eqnarray}
I_{port}&=&qd\int\limits_{port}D_{p}\vec{\nabla}\bar{p}\cdotp\hat{n}dl,
\end{eqnarray}
where $d$ is the thickness of the sample. These diffusive currents must be included in determining the net current
flowing through the device. The applied voltage necessary for maintaining the net current at its given value is
calculated by determining the effective device resistance including all carriers. This voltage is applied as the
boundary condition for the Poisson equation,
\begin{eqnarray}
\vec{\nabla}\cdotp(\varepsilon\vec{\nabla}V)&=&-q\bar{p},
\end{eqnarray}
where $\varepsilon$ is the material permittivity. Equation (15)  is solved using the Galerkin FEM for the electrostatic
potential due to the excess hole density with the boundary conditions that the applied voltage mentioned above is
applied across the current ports 1 and 4, while the voltages at  ports 2 and 3 (see Fig. 1) are determined by the
solution. 

The steps for calculating the potential due to the photo-generated carriers are outlined below for one laser spot
position:
\begin{enumerate}
\item Solve the drift-diffusion equation for $\bar{p}=0$  (Eq. 12).
\item Calculate the current on the input and output current ports (Eq. 14).
\item Determine the applied voltage $V_{14}$ from the requirement that the net current be the given current through the
device.
\item Solve Eq. (15) with $V_{14}$ applied across the current ports.
\item Calculate the voltage difference $V_{23}$ between the voltage ports using the solution of Eq. (15) from the last
step.
\end{enumerate}
\section{\label{sec:RandD}RESULTS AND DISCUSSION}
\subsection{\label{sec:xandy}Dependence of EOC on Laser Spot Position} The measured voltage $V_{23}$ and theoretical
results (solid lines) are shown in Fig. 4 for $y_{\ell}$ = (0.2 mm and 0.7 mm) as a function of $x_{\ell}$  for both the
hybrid and bare samples at a temperature of 15 K with a power density of $6.3\times10^{4}$ W/cm$^{2}$. The theoretical
shape of the hole distribution is governed by the mobility, and thus the mean free path.  The mean free path
 is calculated from 
 \begin{eqnarray}
 l=v_{F}\tau=\left(\frac{h}{m^{*}}\right)k_{F}\left(\frac{\mu m^{*}}{e}\right)=h\sqrt{2\pi
nt}\left(\frac{\mu}{e}\right),
\end{eqnarray}
which takes into account the Fermi velocity, $v_{F}$, Fermi wave vector, $k_{F}$, scattering time, $\tau$, effective
mass, $m^{*}$, mobility, $\mu$,  thickness, $t$, of the sample, and the volumetric carrier density, $n$. As $x_{\ell}$ 
increases there is a peak (valley) corresponding to the positive $V_{2}$ (negative $V_{3}$) voltage lead. As is evident
in Fig. 4, the theoretical model reasonably approximates the positional dependence along the x-direction. The generation
rate $G_{o}$, which was the only adjustable parameter, was varied such that the measured and calculated voltages
coincided at the peak value. The value of the voltage at other $x_{\ell}$ positions was determined using the same
generation rate which was used to match the peak value. This process of choosing the generation rate was done separately
for each value of $y_{\ell}$ and for both the bare and shunted sample. 
\begin{figure}
\includegraphics[width=0.5\textwidth]{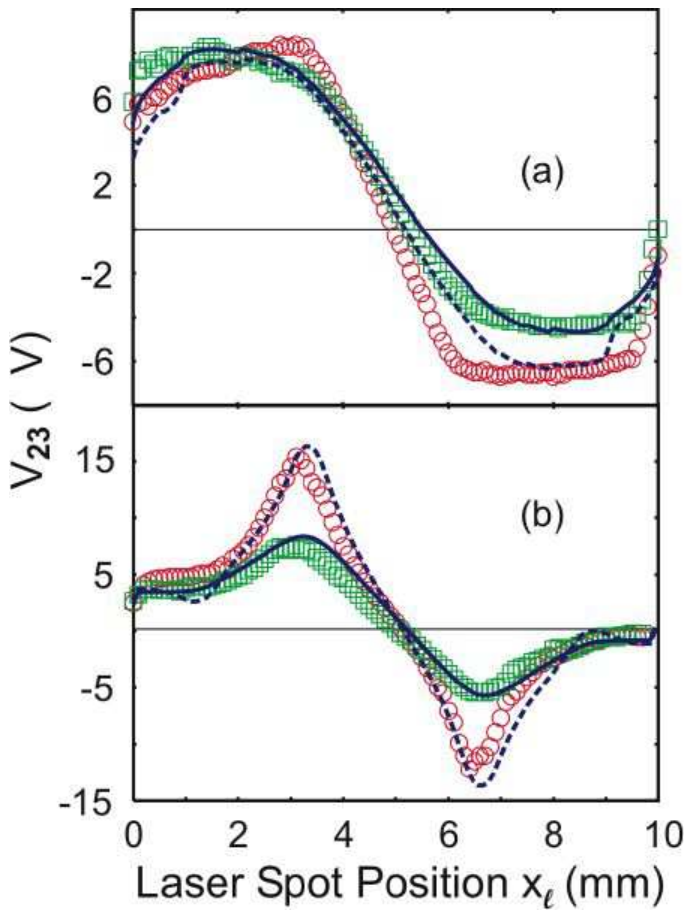}
\caption{\label{fig:V_{x,y}}(Color online) Voltage as the laser spot is moved in the x-direction for $y_{\ell}=0.2$ mm  ($\bigcirc$, - - -) and
$y_{\ell}=0.7$ mm ($\square$,---)
the bare sample (a) and the MSHs (b). The curves represent theoretical calculations and the points represent
experimental measurements.  Experimental measurements were taken at 15 K with a power density of $6.3\times10^{4}$ W/cm$^{2}$.}
\end{figure}

Figure 5 shows the voltage $V_{23}$ for $x_{\ell} = 3.3$ mm as a function of $y_{\ell}$, for both the hybrid and bare
sample with theory represented by solid and dashed lines. As $y_{\ell}$ increases, the absolute value of the voltage
approaches zero for both the hybrid and bare samples.  In addition, with increasing $y_{\ell}$, the charge distributions
are further away from the voltage leads and closer to the shunt, thereby lowering the measured voltage. As a result, the
voltage in the hybrid sample decreases as the laser spot is moved closer to the shunt. Because of these effects, both
$V_{23}$ and the EOC decrease as a function of the $y$ position, with its peak being at $y_{\ell}=0$. Similar to the
$x_{\ell}$ dependence, one generation rate for the $y_{\ell}$ dependence was chosen such that the theoretical voltages
matched the experimental values over the entire range of $y_{\ell}$ values.

\begin{figure}
\includegraphics[width=0.5\textwidth]{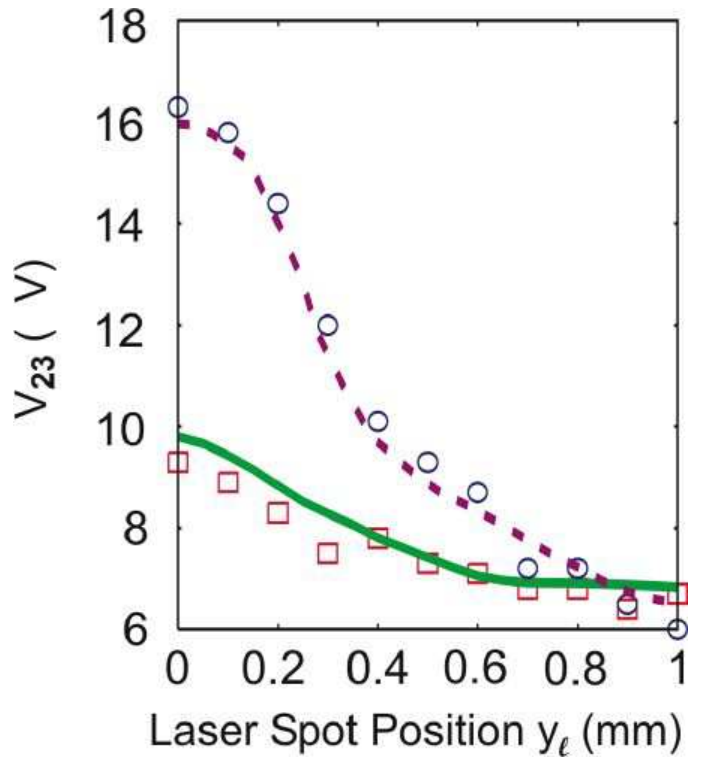}
\caption{\label{fig:V_{y}}(Color online) Voltage as the laser spot is moved in the y-direction for $y_{\ell}=3.3$ mm for the bare
sample ($\Box$,---) and MSHs ($\bigcirc$, - - -). The curves represent theoretical calculations and the points represent
experimental measurements. Relevant experimental parameters are given in Fig. 4 caption.}
\end{figure}

\subsection{\label{sec:EOC(I)}EOC Dependence on the Bias Current} The voltages in both the hybrid and bare samples show
an offset such that when the laser is not illuminating the sample, the measured voltage is nonzero. The voltage offset
was seen to be additive to the voltage produced by the excess carriers over the range of bias current studied. In order
to calculate the EOC, the offset was removed for  both the bare sample and the MSH so that the data reflects the effect
of the excess carriers alone. In addition, the voltage offset was found to be proportional to the bias current which
suggests that it is associated with the intrinsic sample resistance and not the perturbation that we are interested in. 

\subsection{\label{sec:EOC(P))}EOC and Laser Power Dependence} The graph in Fig. 6 shows a plot of the dependence of
$V_{23}$ on the illuminating optical power density P. The theoretical curves indicate that the measured voltage depends
linearly on the generation rate. Similarly, the experimental data indicates that the measured voltage depends linearly
on the optical power density. These two observations suggest that the generation rate is proportional to the incident
power, as should be expected in this regime. For the bare device, the coefficient of proportionality is
$\alpha_{bare}=4\times10^{18}$ J$^{-1}$cm$^{-1}$, and for the shunted device the coefficient is
$\alpha_{shunt}=1\times10^{19}$ J$^{-1}$cm$^{-1}$. The difference in $\alpha$ for the two devices arises from geometric
contributions which affect the generation and recombination of carriers. Note that the range of power spans
approximately two orders of magnitude.

\begin{figure}
\includegraphics[width=0.5\textwidth]{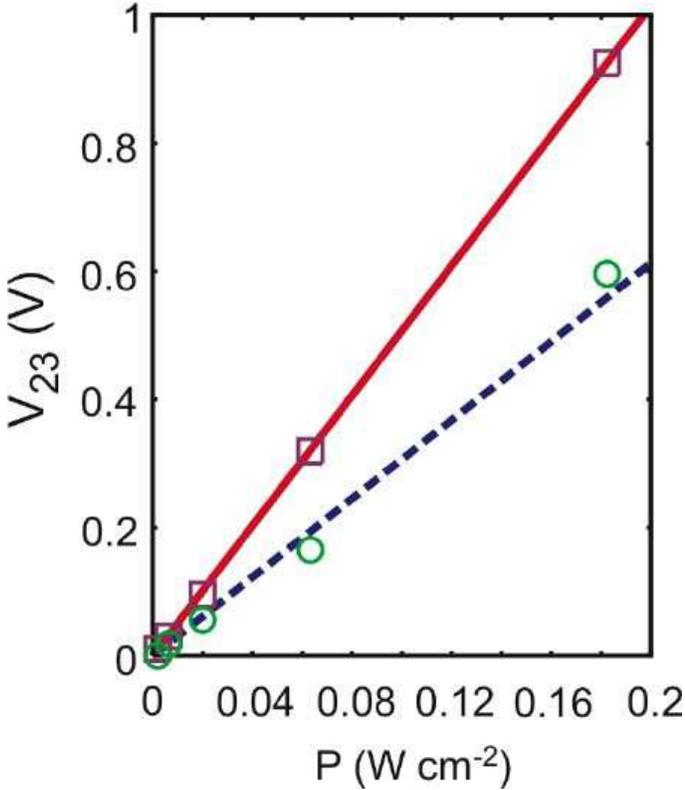}
\caption{\label{fig:V(P))}(Color online) Voltage versus laser power at spot position $(y_{\ell},x_{\ell})=(3.3 $mm$, 0 $mm$)$  for the
bare sample ($\Box$,---) and MSHs ($\bigcirc$, - - -). The lines represent theoretical calculations and the points represent
experimental measurements. The voltage has a linear dependence on the laser power and the generation rate. The
coefficient of proportionality is $\alpha_{bare}=4\times10^{18}$ J$^{-1}$cm$^{-1}$ for the bare device and
$\alpha_{shunt}=1\times10^{19}$ J$^{-1}$cm$^{-1}$ for the shunted device. Experimental measurements were taken at 300 K.}
\end{figure}
\subsection{\label{sec:EOC_{T}}EOC and Temperature Dependence}
 As reported\cite{9} previously, the EOC reaches a maximum of almost 500\% at 30 K for optimal $(x_{\ell},y_{\ell})$
values. The EOC also displays an inverse relationship with temperature.  Here we will quantify and further elucidate
that relationship.

Because the GaAs is degenerately doped, the equilibrium carrier concentrations $(n_{o}, p_{o})$ change minimally over
the temperature range studied.\cite{22} The electron mobility\cite{22} and dielectric constant\cite{16} are also assumed
constant over that temperature range. Implicit in the diffusion and resistive terms in Eq. 3 is a temperature
dependence. According to Lovejoy \textit{et al.}\cite{22} the hole mobility, $\mu_{p}$, has an inverse temperature dependence
that can be fit as $\mu_{p}=\chi T^{-\frac{3}{2}}$ where $\chi$ is a fixed parameter used to fit the data. In Eq. 5, we
see an explicit linear temperature dependence of the diffusion coefficient.
 Thus the combined temperature dependence from Eq. 5 yields $D_{p}\sim T^{-\frac{1}{2}}$. As a result, the effect of the
diffusive component decreases with increasing temperature.

Figure 7 shows the temperature dependence of the EOC for various $y_{\ell}$ values with a power density of
$6.3\times10^{4}$ $W/cm^{2}$ at $\lambda=476.5$ nm and $x_{\ell}=3.3$ mm with the points obtained from experiment and
the curves calculated from the model.  In the model, the generation rate was held constant for the bare GaAs sample,
while $G_{o}$ for the MSH was scaled to fit the EOC value at one temperature. With $G_{o}$ fixed for each $y_{\ell}$ in
this manner, the temperature dependence of the EOC was then calculated by simply incrementing the temperature. This
temperature change, as discussed above, was incorporated directly into the diffusion coefficients as well as the hole
mobility. With the exception of $y_{\ell}=0$ mm and 0.2 mm, the EOC was fitted at 15K.  These two $y_{\ell}$ values were
fitted at 30 K as their 15 K data (shown) was deemed errant owing to the difficulty of establishing the $y_{\ell}=0$
position, as discussed previously.\cite{9} 

\begin{figure}
\includegraphics[width=0.5\textwidth]{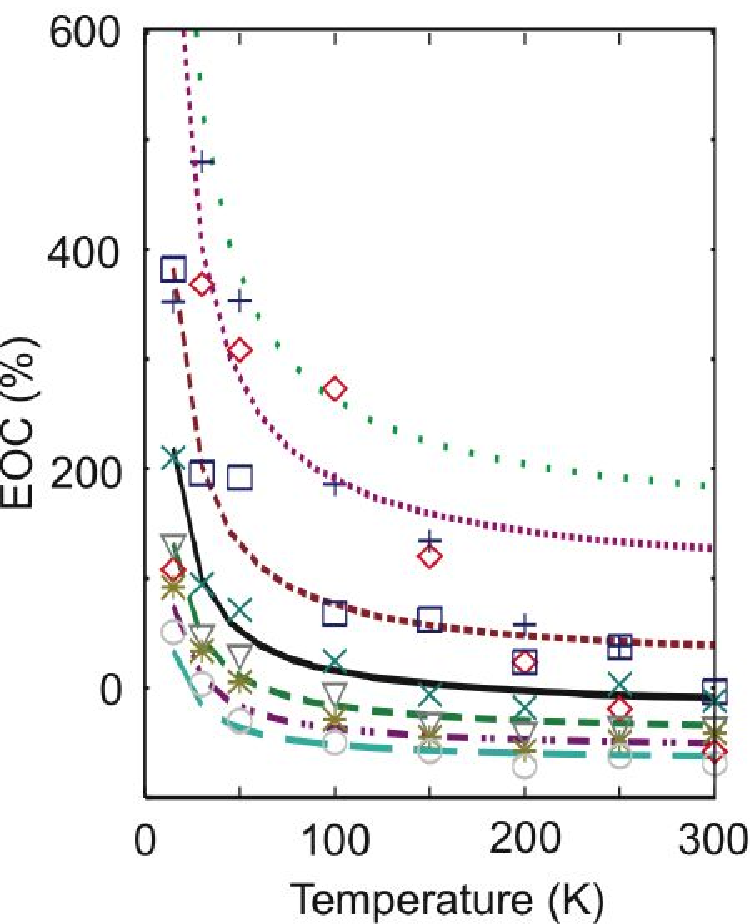}
\caption{\label{fig:V_{T}}(Color online) The temperature dependence of the EOC as defined by Eq. (1).  The measurements were acquired
at spot positions $x_{\ell}=3.3$ mm with 
$y_{\ell}=0$ mm, ($\diamondsuit$, $\cdot$  $\cdot$  $\cdot$);
 $y_{\ell}=0.2$ mm, (\textbf{+}, $\cdots$);
 $y_{\ell}=0.4$ mm, ($\Box$, \textbf{- - -});
 $y_{\ell}=0.6$ mm, ($\times$,solid);
 $y_{\ell}=0.8$ mm, ($\triangledown$, $\relbar\relbar\relbar$)    
 $y_{\ell}=1.0$ mm, (\textbf{$\ast$},$\relbar$ $\cdot\cdot$ $\relbar$); and
 $y_{\ell}=1.2$ mm, ($\bigcirc$, $\relbar$ $\relbar$ $\relbar$)    
Experimental measurements were taken with a power density of $6.3\times10^{4}$ W/cm$^{2}$.}
\end{figure}

\subsection{\label{sec:eocvlpe}EOC and the lateral photovoltaic effect} Because of the similiarity in experimental setup
and positional response to laser illumination between EOC and the well known lateral photovoltaic effect (LPE),\cite{23}
a discussion highlighting the differences is appropriate. Critical to the discussion of EOC and LPE is the illumination
of the junction. The lateral photovoltaic effect requires illumination of the junction between a (doped) semiconductor
and a metal or a heavily doped (conducting) semiconductor, using a configuration in which the illuminating beam is
incident normal to the plane of the junction.\cite{24}  In contrast, in the EOC effect, the junction is NOT illuminated
by the incident laser beam, which is incident parallel to the plane of the junction. Also, the LPE relies on a Schottky
barrier at the metal-semiconductor interface, while for the EOC devices shown here, the interface is an ohmic contact.

Beginning with Wallmark\cite{23} all theoretical analyses of the LPE are based on carrier dynamics derived from the
photocurrent generated across the junction whereas our analysis, as explained above, is based primarily on the Dember
effect and provides a very good fit to the position dependent voltage (see Figs. 4 \& 5). Thus, despite the similarity
of the positional dependence of the voltage response seen in some LPE measurements and ours, it should be clear that the
EOC is a different phenomenon.

In the LPE, the output voltage at any lateral position of the illumination INCREASES with decreasing
temperature.\cite{24,25} In the EOC effect in GaAs-In, the output voltage DECREASES with decreasing temperature. Our
theoretical model provides a good fit to the temperature dependence of the EOC (see Fig. 7). This further supports the
fact that EOC and LPE are fundamentally different position dependent effects.
\section{\label{sec:conclusion}CONCLUSION}

The drift diffusion model of EOC has been successfully employed to fit experimental data taken using GaAs-In MSH
structures.  The model based on the FEM closely fits the voltage data and the resulting EOC calculation.  The position
dependence in x and y is incorporated in the model via the drift diffusion equation for holes based on the mean free
path, and thus the mobility.  By including a generation and recombination rate a reasonable fit was obtained. The model
also reproduced the decrease in EOC with increasing $y_{\ell}$. The EOC decreases with increasing temperature $T$, which
the model directly accounts for through the temperature dependence of the mobility and the diffusion coefficients. 

As mentioned briefly in our preliminary report\cite{9} EOC is similar to but fundamentally different from the well known
lateral photovoltaic effect (LPE). The main difference lies in the fact that the interface is not illuminated in EOC,
whereas in the LPE such illumination is critical. Moreover, because of the many factors influencing the EOC, it presents
a unique opportunity to study and model the dynamics of the metal-semiconductor interface, carrier transport phenomena,
and the effect of introducing a shunt.

Because the carrier mobilities are critical in the Dember effect, InSb should yield a two orders of magnitude increase
in the observed EOC, possibly even room temperature EOC, as compared to GaAs. Using the drift diffusion model, we
anticipate modeling the EOC of this direct gap semiconductor and others. We have not experimentally or theoretically
optimized for wavelength, geometry of the device, or material parameters. With the new ability to model EOC as shown,
predictions can be made to maximize the gain of an EOC device.  Once optimized, we may expect MSH devices based on EOC
to impact optical sensing technology, particularly position sensitive detectors.

\begin{acknowledgments} The authors would like to thank Professor S. E. Hayes and Professor D. L. Rode for providing and
dicing the GaAs samples.  We also thank Yue Shao for useful discussions. This work is supported by the US National
Science Foundation under grant ECS-0329347.  LRR thanks Washington University in St. Louis for the Clay Way Harrison Fellowship which supported his collaborative visit during part of which this work was carried out.
\end{acknowledgments}


\end{document}